\documentclass[11pt]{article}

\newcommand{\be}{\begin{equation}}
\newcommand{\ee}{\end{equation}}
\def\bea{\begin{eqnarray}}
\def\eea{\end{eqnarray}}
\def\der{\partial}

\newcommand{\mscr}[1]{\mbox{\scriptsize #1}}

\newcommand{\ft}[2]{{\textstyle\frac{#1}{#2}}}

\newcommand{\Smacro}{{\mathcal S}_{\mscr{macro}}}
\newcommand{\Smicro}{{\mathcal S}_{\mscr{micro}}}

\usepackage{amsmath}
\usepackage{amssymb}
\usepackage{amscd}
\usepackage{bbm}

\begin{document}

\renewcommand{\thefootnote}{\fnsymbol{footnote}}

\begin{titlepage}
\begin{center}
\hfill LTH 741 \\
\hfill {\tt hep-th/0703037}\\
\vskip 10mm

{\Large
{\bf Special Geometry, Black Holes and Euclidean
Supersymmetry}\footnote{Contribution to the Handbook on 
Pseudo-Riemannian Geometry and Supersymmetry, ed. V. Cort\'es,
published by the European Mathematical Society in the series
``IRMA Lectures in Mathematical and Theoretical Physics.''}
}

\vskip 10mm

\textbf{Thomas Mohaupt} 

\vskip 4mm

Theoretical Physics Division\\
Department of Mathematical Sciences\\
University of Liverpool\\
Liverpool L69 3BX, UK \\
{\tt Thomas.Mohaupt@liv.ac.uk}
\end{center}

\vskip .2in 

\begin{center} {\bf ABSTRACT} \end{center}
\begin{quotation} \noindent
We review recent developments in special geometry and
explain its role in the theory of supersymmetric black holes.
To make this article self-contained, a short introduction
to black holes is given, with emphasis on the laws
of black hole mechanics and black hole entropy. We also
summarize  the existing results on the 
para-complex version of special geometry, which 
occurs in Euclidean supersymmetry. The role of real coordinates
in special geometry is illustrated, and we briefly 
indicate how Euclidean supersymmetry can be used to 
study stationary black hole solutions via dimensional
reduction over time. \\
This article is an updated and substantially extended version of the previous 
review article `New developments in special geometry', hep-th/0602171.
\end{quotation}

\vfill

\end{titlepage}

\eject

\renewcommand{\thefootnote}{\arabic{footnote}}

\section{Introduction}
\setcounter{equation}{0}
\setcounter{footnote}{0}

Special geometry was discovered more than 20 years ago \cite{deWvP}. While 
the term special geometry originally referred to the geometry of vector 
multiplet
scalars in four-dimensional $N=2$ supergravity, today it is used 
more generally for the geometries encoding the scalar couplings of 
vector and hypermultiplets in theories with 8 real supercharges. 
It applies to rigidly and locally supersymmetric theories in $\leq 6$ 
space-time
dimensions, both in Lorentzian and in Euclidean signature. 
The scalar geometries occuring in these cases are indeed
closely related. In particular, they are all much more restricted
than the K\"ahler geometry of scalars in theories with 4 supercharges,
while still depending on arbitrary functions.
In contrast, the scalar geometries of theories with 16 or
more supercharges are completely fixed by their matter content. 
Theories with 8 supercharges have a rich dynamics, which is
still constrained enough to allow one to answer many questions exactly.
Special geometry lies at the heart of the Seiberg-Witten solution
of $N=2$ gauge theories \cite{SW}
and of the non-perturbative dualities
between $N=2$ string compactifications \cite{KV,2nd}.

While the subject has now been studied for more than twenty years,
there are still new aspects to be discovered. One, which will be
the topic of this article, is the role of real coordinates. Many special
geometries, in particular the special K\"ahler manifolds of 
four-dimensional vector multiplets and the hyper-K\"ahler geometries
of rigid hypermultiplets are complex
geometries. Nevertheless, they also possess distinguished real
parametrizations, which are natural to use for certain physical
problems. Our first example illustrates this in the context of 
special geometries in theories with  Euclidean supersymmetry. This part
reviews the results of \cite{CMMS1,CMMS2}, and gives us the
opportunity to explore another less studied aspect of special 
geometry, namely the scalar geometries of $N=2$ supersymmetric
theories in Euclidean space-time.
It turns out that the relation 
between the scalar geometries of theories with Lorentzian and 
Euclidean space-time geometry is (roughly) given by replacing complex 
structures by para-complex structures. One technique for deriving
the scalar geometry of a Euclidean theory in $D$ dimensions is to
start with a Lorentzian theory in $D+1$ dimensions and to perform 
a dimensional reduction along the time-like direction. The specific
example we will review is to start with vector multiplets in 
four Lorentzian dimensions, which gives, by reduction over time,
hypermultiplets in three Euclidean dimensions. This provides us with
a Euclidean version of the so-called c-map. The original c-map
\cite{CFS,FS}
maps any scalar manifold of four-dimensional vector multiplet
scalars to a scalar manifold of hypermultiplets. For rigid 
supersymmetry, this relates affine special K\"ahler manifolds
to hyper-K\"ahler manifolds, while for local supersymmetry this
relates projective special K\"ahler manifolds to quaternion-K\"ahler 
manifolds. By using dimensional reduction with respect to
time rather than space, we will derive the scalar geometry of
Euclidean hypermultiplets. As we will see,
the underlying geometry is particularly transparent 
when using real  scalar fields rather than complex ones. 
The geometries of Euclidean supermultiplets are relevant for
the study of instantons, 
and, by `dimensional oxidation over time' also for solitons, as
outlined in \cite{CMMS1}. In this article we will restrict
ourselves to the geometrical aspects.

Our second example is taken from a different context, namely BPS black hole
solutions of matter-coupled $N=2$ supergravity. The laws of black hole
mechanics suggest to assign an entropy to black holes, which is, at least
to leading order, proportional to the area of the event horizon. 
Since (super-)gravity presumably is the low-energy effective theory of
an underlying quantum theory of gravity, the black hole entropy is
analogous to the macroscopic or thermodynamic entropy in thermodynamics.
A quantum theory of gravity should provide the fundamental or microscopic
level of description of a black hole and, in particular, should allow
one to identify the microstates of a black hole and to compute 
the corresponding microscopic or statistical entropy. The 
microscopic entropy is the information missing if one only knows
the macrostate but not the microstate of the black hole. In other
words, if a black hole with given mass, charge(s) and angular 
momentum (which characterise the macrostate) can be in $d$
different microstates, then the microscopic entropy is
$S_{\mscr{micro}} = \log d$. If the area of the event horizon
really is the corresponding macroscopic entropy, then these
two quantities must be equal, at least 
to leading order in
the semi-classical limit. In string theory it has been shown 
that the two entropies are indeed equal in this limit
\cite{StrVaf},
at least for BPS states (also called supersymmetric
states). These are states 
which sit in special representations of the supersymmetry algebra,
where part of the generators act trivially. These BPS (also called short)
representations saturate the lower bound set for the mass by
the supersymmetry algebra, and, as a consequence, the mass is exactly
equal to a central charge of the algebra.\footnote{See \cite{WaB}
Chapter 2.}  In this article we will
be interested in the macroscopic part of the story, which is the
construction of BPS black hole solutions and the computation of 
their entropy. The near horizon limit of such solutions, which is 
all one needs to know in order to compute the entropy, is determined
by the so-called black hole attractor equations \cite{AttractorEqs}, 
whose derivation
is based on the special geometry of vector multiplets. The
attractor equations are another example where real coordinates on the
scalar manifold appear in a natural way. In the second part of the 
article we review how the attractor equations and the entropy can 
be obtained from a variational principle. When expressed in terms
of real coordinates, the variational principle states that the
black hole entropy is the Legendre transform of the Hesse potential 
of the scalar manifold. We also discuss how the black hole free
energy introduced by Ooguri, Strominger and Vafa \cite{OSV} fits into 
the picture, 
and indicate how higher curvature and non-holomorphic corrections
to the effective action can be incorporated naturally. This 
part of the article is based on \cite{CdWKM:06} and on older work
including \cite{BCdWKLM, CdWM:98, CdWKM:00}.

Let us now explain 
how our two subjects are connected to the second topic of this volume,
pseudo-Riemannian geometry. In both parts of the article
we have  two relevant geometries, the geometry of space-time and the
geometry of the target manifold of the scalar fields. In the first
case, space-time is Euclidean, but, as we will see,
the scalar manifold is pseudo-Riemannian with split signature. 
In the second case the scalar geometry is positive definite, 
but space-time is pseudo-Riemannian with Lorentz signature. 
In fact, our two subjects, the c-map and black holes, can be
related in a rather direct fashion, as follows: for a static
black hole one can perform a dimensional reduction along the
time-like direction in complete analogy to the dimensional reduction 
of flat Minkowski space. Then one can dualize the vector multiplets
into hypermultiplets, which gives rise to a `local' version of 
the c-map.\footnote{Here local means that supersymmetry is realized
as a local, i.e., space-time dependent symmetry.} 
This construction can be used to study 
time-independent four-dimensional geometries from a three
dimensional perspective, which has the advantage that 
all bosonic degrees (metric, gauge fields and scalars)
become scalars in the reduced theory and 
can then be combined into a non-linear sigma model.
This method has been used in Einstein-Maxwell theory
already a long time \cite{NK}, and has been elaborated 
on both for black holes \cite{BMG} and brane-type solutions \cite{CGR}.
We refer to \cite{Stelle} for a review. 
More recently, dimensional reduction to three Euclidean dimensions 
and the corresponding version of the c-map have been used 
by \cite{GNPW} to elaborate on the ideas of Ooguri, Strominger
and Vafa \cite{OSV} by quantizing static, spherically BPS black hole 
solutions.
The three-dimensional solutions obtained by dimensional reduction
for four-dimensional static black hole solutions can also 
be lifted to four Euclidean dimensions, where they
desribe a wormhole solutions, 
which generalize the D-instanton of type-IIB string theory 
\cite{BGLMM}. 

Let us finally mention two contributions to this volume which 
are closely related to our topics. The article \cite{RVV2}
(which is based on \cite{RVV}) disucsses new insights into the
geometry of the c-map, which have been obtained by relating
vector and hypermultiplets to tensor multiplets. The contribution
of \cite{IMZ} discusses new developments in para-quaternionic 
geometry. While we only discuss the `rigid' version of the Euclidean 
c-map in this article, its `local' (supergravity) version maps
projective special K\"ahler manifolds to para-Quaternionic manifolds.

\section{Euclidean special geometry}

\subsection{Vector multiplets}

We start by reviewing the geometry of vector multiplets in 
rigid four-dimen\-sional $N=2$ supersymmetry.\footnote{Some more
background material and references on vector multiplets can be
found in \cite{Habil}.} A vector multiplet consists
of a gauge field $A_m$, ($m=0,\ldots, 3$ is the Lorentz index),
two Majorana spinors $\lambda^i$ ($i=1,2$) and
one complex scalar $X$. We consider $n$ such multiplets, labeled
by an index $I=1, \ldots, n$. The field equations for the gauge
fields are invariant under $Sp(2n, \mathbbm{R})$ rotations which act linearly
on the field strength $F^I_{m n}$ and the dual field
strength $G_{I | m n} = \ft{\delta L}{\delta F^I_{m n}}$,
where $L$ denotes the Lagrangian. These symplectic rotations
generalize the electric-magnetic duality rotations of Maxwell
theory and are in fact invariances of the full field equations. 
A thorough analysis shows that this has the important consequence
that all vector multiplet couplings are encoded in a single holomorphic 
function of  the scalars, $F(X^I)$, which is called the 
prepotential \cite{deWvP}.  In 
superspace language the general action for vector multiplets can
be written as a chiral superspace integral of the prepotential 
$F$, considered as a superspace function of  
$n$ so-called restricted chiral multiplets $(X^I, \lambda^{I+}, 
F^{I-}_{m n})$, which encode the gauge invariant quantities 
of the $n$ vector multiplets. Here $\lambda^{I+}$ are the positive
chirality projections of the spinors and $F^{I-}_{m n}$ are
the antiselfdual projections of the field strength. To be precise, the
Lagrangian is the sum of a chiral and an antichiral superspace integral,
the latter depending on the complex conjugated multiplets
$(\overline{X}^I, \lambda^{I-}, F^{I+}_{m n})$. When working out
the Lagrangian in components, all couplings can be expressed in 
terms of $F(X^I)$, its derivatives, which we denote $F_I, F_{IJ}, \ldots$
and their complex conjugates $\overline{F}_I, \overline{F}_{IJ}, \ldots$. 
For later use we specify the bosonic
part of the Lagrangian:
\be 
L_{\mscr{bos}}^{\mscr{4d VM}} = - \ft12 N_{IJ} \der_m X^I \der^m
\overline{X}^J - \ft{i}{2}  \left( F_{IJ} F^{I-}_{m n} F^{J-mn}
- \mbox{c.c.} \right) \;,
\label{4dVMlagrangian}
\ee
where
\be
N_{IJ} = \der_I \der_{\overline{J}} 
\left(-i (X^I \overline{F}_I - F_I \overline{X}^I)
\right)
\label{KaePotLor}
\ee
can be interpreted as a Riemannian metric on the target space $M_{VM}$
of the scalars $X^I$.\footnote{The scalar fields $X^I$ might
only provide local coordinates. We will work in a single coordinate
patch  throughout.}
$N=1$ supersymmetry requires this metric to be a K\"ahler
metric, which is obviously the case, the K\"ahler potential being
$K = -i (X^I \overline{F}_I - F_I \overline{X}^I)$. As a consequence
of $N=2$ supersymmetry the metric is not a generic K\"ahler metric,
since the K\"ahler potential can be expressed in terms of the
holomorphic prepotential $F(X^I)$. The resulting geometry is known as
affine (also: rigid) special K\"ahler geometry. The intrinsic characterization
of this geometry is the existence of a flat, torsionfree, symplectic
connection $\nabla$, called the special connection, such that 
\be
(\nabla_U I) V = (\nabla_V I) U \;,
\ee
where $I$ is the complex structure and $U, V$ are arbitrary vector 
fields \cite{Freed}. 
It has been shown that all such manifolds can be constructed
locally as holomorphic Langrangian immersions into the complex symplectic
vector space $T^* {\mathbbm C}^n \simeq \mathbbm{C}^{2n}$ \cite{ACD}. In this
context $X^I, F_I$ are flat complex symplectic coordinates on $T^* 
\mathbbm{C}^n$ and the prepotential is the generating function of
the immersion $\Phi: M_{VM} \rightarrow T^* \mathbbm{C}^n$, i.e., 
$\Phi  = d F$. For generic choice of $\Phi$, the $X^I$ provide 
coordinates on the immersed $M_{VM}$, while 
$F_I = \der_I F= F_I(X)$ along $M_{VM}$ . 
The $X^I$ are non-generic coordinates, physically, because they are the
lowest components of vector multiplets, mathematically, 
because they are adapted to the immersion. They are called special
coordinates.

So far we have considered vector multiplets in a four-dimensional
Min\-kows\-ki space-time. In four-dimensional Euclidean space the 
theory has the same form, except that the complex structure 
$I$, $I^2 = - \mathbbm{1}$ is replaced by a para-complex structure
$J$. This is an endomorphism of $T M_{VM}$ such that 
$J^2 = \mathbbm{1}$, with the eigendistributions corresponding
to the eigenvalues $\pm 1$ having equal rank. Many notions of 
complex geometry, including K\"ahler and special K\"ahler geometry
can be adapted to the para-complex realm. We refer to \cite{CMMS1,
CMMS2} for the details. In particular, it can be
shown that the target space geometry of rigid Euclidean vector 
multiplets is affine special para-K\"ahler. Such manifolds
are the para-complex analogues of affine special K\"ahler manifolds.
When using an appropriate
notation, the expressions for the Lagrangian, the equations of motion 
and the supersymmetry transformation rules take the same form
as for Lorentzian supersymmetry, except that complex quantities
have to be re-interpreted as para-complex ones. For example, 
the analogue of complex coordinates $X^I = x^I + i u^I$, where
$x^I, u^I$ are real and $i$ is the imaginary unit, are para-complex
coordinates $X^I = x^I + e u^I$, where $e$ is the para-complex
unit characterized by $e^2=1$ and $\overline{e} = - e$, where 
the `bar' denotes para-complex conjugation.\footnote{It has been
known for quite a while that the Euclidean version of a supersymmetric
theory can sometimes be obtained by replacing $i \rightarrow e$
\cite{GGP}.}
While in Lorentzian
signature the selfdual and antiselfdual projections of the 
field strength are related by complex conjugation, in the Euclidean
theory one can re-define the selfdual and antiselfdual projections 
by appropriate factors of $e$
such that they are related by para-complex conjugation. One can also
define para-complex spinor fields such that the fermionic terms
of the Euclidean theory take the same form as in the Lorentzian one.
The Euclidean bosonic Lagrangian takes the same form (\ref{4dVMlagrangian})
as the Lorentzian one, with (\ref{KaePotLor}) replaced by
\be
N_{IJ} = \der_I \der_{\overline{J}} 
\left(-e (X^I \overline{F}_I - F_I \overline{X}^I)
\right) \;.
\label{KaePotEuc}
\ee
Note that the Euclidean Lagrangian is real-valued, although the
fields $X^I$ and $F^{I-}_{mn}$ are para-complex. We also remark
that a para-K\"ahler metric always has split signature. 
The full Lagrangian, including fermionic terms, and the
supersymmetry transformation rules can be found in \cite{CMMS1}.
There we also verified that it is related to the rigid limit
of the general Lorentzian signature vector multiplet
Lagrangian \cite{deWLauVanP,DeJdeWKV}
by replacing $i \rightarrow e$ (together with
additional field redefinitions, which account
for different normalizations and conventions).

\subsection{Hypermultiplets} 

Our next step is to derive the geometry of Euclidean 
hypermultiplets. This can be done by either reducing the Lorentzian
vector multiplet Lagrangian with respect to time or the Euclidean
vector multiplet Lagrangian with respect to space \cite{CMMS2}. 
Here we start from the Lorentzian Lagrangian and perform 
the reduction over space and  over time in parallel.
This is instructive, because
the reduction over space corresponds to the standard 
$c$-map and gives us hypermultiplets in three-dimensional Minkowski space-time,
while the reduction over time is the new para-$c$-map and gives 
us hypermultiplets in three-dimensional Euclidean space.

Before performing the reduction, we rewrite the Lorentzian 
vector multiplet Lagrangian in terms of real fields. Above we noted
that the intrinsic characteristic of an affine special K\"ahler 
manifold is the existence of the special connection $\nabla$, which 
is, in particular, flat, torsionfree and symplectic \cite{Freed}. 
The corresponding
flat symplectic coordinates are 
\be
x^I = {\mbox Re} X^I \;\;\;
y_I = {\mbox Re} F_I  \;.
\ee
Note that since $F$ is an arbitrary holomorphic function, these
real coordinates are related in a complicated way to the special
coordinates $X^I$. The real coordinates $x^I, y_I$ are flat (or 
affine) coordinates with respect to $\nabla$, i.e., 
$\nabla dx^I = 0 = \nabla dy_I$, and they are symplectic (or 
Darboux coordinates), because the symplectic form on $M_{VM}$ 
is $\omega = 2 dx^I \wedge dy_I$. While in special coordinates
the metric of $M_{VM}$ can be expressed in terms of the prepotential
by (\ref{KaePotLor}), the metric has a Hesse potential when using
the real coordinates $q^a = (x^I, y_I)$, where
$a=1, \ldots, 2n$ \cite{Freed,Hitchin}:
\be
g_{ab} = \frac{ \der^2 H }{\der q^a \der q^b} \;.
\ee
The Hesse potential is related to the imaginary part of the 
prepotential by a Legendre transform \cite{Cor}:
\be
H(x,y) = 2 {\rm Im} F(x+iu) - 2 u^I y_I  \;.
\label{HessePrepot}
\ee
The two parametrizations of the metric on $M_{VM}$ are related by
\be
ds^2 = - \ft12 N_{IJ} d X^I d \overline{X}^J = - g_{ab} dq^a dq^b \;.
\ee
In order to rewrite the Lagrangian (\ref{4dVMlagrangian}) completely
in terms of real fields, we express the (anti)selfdual field strength
$F^{I\pm}_{mn}$ in terms of the field strength $F^I_{mn} =
F^{I+}_{mn} + F^{I-}_{mn}$ and their Hodge-duals 
$\tilde{F}^{I}_{mn} =i( F^{I+}_{mn}-
F^{I -}_{mn})$. The result is
\be
\label{4dVMlagrReal}
L_{\mscr{bos}}^{\mscr{4d VM}} = - g_{ab} \der_m q^a \der^m q^b 
- \ft14 N_{IJ} F^I_{mn} F^{Jmn}
+ \ft14 R_{IJ} F^I_{mn} \tilde{F}^{Jmn} \;,
\ee
where 
\bea
R_{IJ} &=& F_{IJ} + \overline{F}_{IJ} \;,\nonumber \\
N_{IJ} &=& i(F_{IJ} - \overline{F}_{IJ}) = \der_I \der_{\overline{J}}
\left( -i (X^I \overline{F}_{I} - F_I \overline{X}^I ) \right) \;.
\eea

We now perform the reduction of the Lagrangian (\ref{4dVMlagrReal})
from four to three dimensions. We treat the reduction over space and
over time in parallel. In the following formulae, $\epsilon=1$ refers
to reduction over time, which gives a Euclidean three-dimensional theory,
while $\epsilon=-1$ refers to reduction over space. By reduction,
one component of each gauge field becomes a scalar. We define:
\be
p^I = A^{I|0} \mbox{   for   } \epsilon = 1 \;,\;\;\;\;
p^I = A^{I|3} \mbox{   for   } \epsilon = -1 \;.
\ee
Moreover, the $n$ three-dimensional gauge fields $A^{I| \hat{m}}$
obtained from dimensional reduction\footnote{
The three-dimensional vector index takes values
$\hat{m}= 0,1,2$ for $\epsilon =-1$ and
$\hat{m}= 1,2,3$ for $\epsilon =1$.} 
can be dualized into $n$ 
further real scalars $s_I$. Denoting the new scalars by 
\be
(\hat{q}_a) = (s_I, 2 p^I) \;,
\ee
the reduced bosonic Lagrangian takes the following, remarkably 
simple form:
\be
L_{HM} = - g_{ab}(q) \der_i q^a \der^i q_b + \epsilon g^{ab}(q) \der_i 
\hat{q}_a \der^i \hat{q}_b \;,
\label{HMlagr}
\ee
where $g^{ab}(q)$ is the inverse of $g_{ab}(q)$.
In this parametrization it is manifest  that the hypermultiplet target
space with metric $(g_{ab}(q)) \oplus (- \epsilon g^{ab}(q))$ is 
$N=M_{HM} = T^* M_{VM}$. The geometry underlying this Lagrangian
was presented in detail in \cite{CMMS2} for $\epsilon = 1$, 
and works analogously for $\epsilon=-1$. 
Here we give a brief summary. The special connection $\nabla$ on
$M=M_{VM}$, can be used to define a decomposition
\be
T_{\xi} N = {\cal H}^\nabla_\xi \oplus T_\xi^v N  
\simeq T_q M \oplus T^*_q M \;,
\ee
where $\xi \in N$ is a point on $N$ (with local
coordinates $(q^a, \hat{q}_a)$), $q= \pi(\xi) \in M$ is its
projection onto $M$,  
${\cal H}^\nabla_\xi$ is the horizontal subspace with respect 
to the connection $\nabla$ and $T_\xi^v N$ is the vertical subspace.
The identification with $T_q M \oplus T^*_q M$ is canonical,
and the scalar fields $q^a, \hat{q}_a$ obtained by dimensional
reduction are adapted to the decomposition. One can then 
define a complex structure $J_1$ on $N$, which acts on 
$T_\xi N \simeq T_qM \oplus T_q^*M$ by multiplication with
\be
J_1 := J_1^\nabla = \left( \begin{array}{cc}
J & 0 \\ 0 & J^*
\end{array} \right) \;,
\ee
where $J$, $J^*$ denote the action of the complex structure
$J$ of $M$ on $TM$ and $T^*M$, respectively. Let us now 
consider
the Euclidean case $\epsilon =1$ for definiteness.
Using the K\"ahler form
$\omega$ on $M$, one can further define 
\be
J_2  = \left( \begin{array}{cc}
0 & \omega^{-1} \\
\omega & 0 \\
\end{array} \right) \;,
\ee
where $\omega$ is interpreted as a map $T_q M \rightarrow T^*_q M$.
This is a para-complex structure, 
$J_2^2 = \mathbbm{1}$. Moreover, $J_3=J_1 J_2 $ is a second 
para-complex structure, and $J_1,J_2,J_3$ satisfy a 
modified version of the quaternionic algebra known as
the para-quaternionic algebra. Thus, $(J_1, J_2, J_3)$ 
is a para-hyper-complex structure on $N$. When defining,
as in (\ref{HMlagr}),
the metric on $N$ by
\be
g_N = \left( \begin{array}{cc}
g & 0 \\ 0 & -g^{-1} \\
\end{array} \right) \;,
\ee
where $g$ is the metric on $M$, then $J_1$ is an isometry,
while $J_2,J_3$ are anti-isometries. This means that
$(J_1, J_2, J_3,g_N)$ is a para-hyper Hermitian structure.\footnote{
Also note that $J_1,J_2,J_3$ are integrable, which follows
from the integrability of $J$.}
Moreover, the structures $J_\alpha$, $\alpha=1,2,3$  are
parallel with respect to the Levi-Civita
connection on $N$. Thus the metric $g_N$ is para-hyper K\"ahler,
meaning that it is K\"ahler with respect to $J_1$ and
para-K\"ahler with respect to $J_2, J_3$. The case 
$\epsilon=-1$ works analogously. Here one finds
three complex structures satisfying the quaternionic algebra,
and the metric defined by (\ref{HMlagr}) is hyper-K\"ahler.

One can introduce (para-)complex fields such that one
of the complex or (para-)complex structures becomes manifest
in the three-dimensional Lagrangian \cite{CFS,CMMS2}. 
In these coordinates the Lagrangian is more complicated, and
the geometrical structure reviewed above is less clear. 
Moreover one has singled out one of the three 
(para-)complex structures. Thus working in real coordinates
has advantages, which should be exploited further in the
future. Note in particular that for the c-map in local
supersymmetry, the target space of hypermultiplets is
quaternion-K\"ahler for Lorentzian space-time, while it
is expected to be para-quaternion-K\"ahler for Euclidean
space-time. In general, the almost (para-)complex
structures of a (para-)quaternion-K\"ahler manifold 
need not be integrable.  Then
combining real scalar fields into (para-)complex
fields is not natural, as these fields do not
define local (para-)complex coordinates.

\section{Black holes  \label{BHsandStrings}}

In order to prepare for our second application of
special geometry, we now give a brief self-contained
introduction to certain aspects of black holes.\footnote{
See \cite{Wald1,Wald2,FroNov} for a detailed discussion.} 
Somewhat surprisingly, one can associate thermodynamic
properties to black holes. The so called laws of 
black hole mechanics, which have been derived in the
framework of classical, matter-coupled Einstein gravity,
formally have the same structure as the laws of 
thermodynamics \cite{BCH}.
While this was originally suspected to be a coincidence,
the (theoretical) discovery of the Hawking effect \cite{Haw} 
strongly 
suggested to take this observation 
seriously. More recently, developments in string theory
have provided additional insights. Let us now briefly
review this, starting with the laws of black hole
mechanics in classical gravity.

\subsection{The laws of black hole mechanics}

The zeroth laws of black hole mechanics states that
the surface gravity $\kappa_S$ of a black hole 
is constant over the event horizon $\Delta$.\footnote{The
term `horizon' is unfortunately used for two different but
closely related concepts. We will use $\Delta$ to denote
the null hypersurface which is the boundary between the
exterior and interior of the black hole in space-time,
and $H$ for the space-like surface which is the boundary
in space, at given time. Thus $H$ is a spatial section 
of $\Delta$ while $\Delta$ is the `worldline' of $H$.} 
The surface 
gravity can be defined if the event horizon is a 
Killing horizon, which is the case for all stationary
black hole solutions of matter-coupled Einstein gravity. 
A Killing horizon is a hypersurface in space-time where
a Killing vector field $\xi$ becomes null:
$\xi^\nu \xi_\nu = 0$. One can show
that a Killing horizon is generated by the integral lines
of the Killing vector fields, which are null geodesics.
There are two natural normal vector fields: 
the Killing vector field $\xi$ itself and the gradient 
of its normed-squared, $\nabla(\xi^\nu \xi_\nu)$.
Both vector fields must be 
proportional, and the factor of proportionality
is defined to be the surface gravity:
\be
\nabla^\mu ( \xi^\nu \xi_\nu) = - 2 \kappa_S \xi^\mu \;.
\ee
While this implies that $\kappa_S$ is a function on the 
horizon, the zeroth law states that this function is
constant. The physical interpretation of the surface
gravity is that it measures the force which an observer
outside the black hole must apply in order to keep a 
unit test mass fixed at the horizon. Thus it measures the
strength of gravity at the horizon. Since the zeroth law
of thermodynamics is that temperature is constant in 
thermodynamical equilibrium, this suggests to interprete
surface gravity as temperature and stationary black holes
as equilibrium states. At the classical level this 
interpretation cannot be defended against the obvious
problem that a black hole does not emit radiation,
a fact 
which is explicitly alluded to in the term `black' hole.
As we will review below this changes once quantum
effects are taken into account. For the time being we
focus on the assumptions needed to prove the zeroth law. 
The classical proof
uses the explicit form of the Einstein equations, while 
the effects of matter are controlled by imposing a
suitable condition on the energy-momentum tensor. Moreover
the solution must be stationary. It then follows from
the field equations that the event horizon is a Killing
horizon. However, it was realized later that the zeroth 
law does not depend on the details of the gravitational
field equations. Instead, it can be proved for any 
covariant (diffeomorphism invariant) action, including 
actions which contain higher derivative and in particular 
higher curvature terms \cite{RacWal}. The relevance of such
actions will be discussed below. The prize for not
specifying the field equations is that one needs
to make the following assumptions: (i) the field
equations admit stationary black hole solutions with
a Cauchy hypersurface, (ii)
the event horizon is a Killing horizon, (iii) 
if the black hole is stationary but not static, then
certain symmetry properties, which in Einstein geometry
are consequences of the field equations, need to be
imposed.\footnote{A space-time is stationary if it 
admits a time-like Killing vector field. In a static
space-time this Killing vector field is in addition
required to be the normal vector field of a family 
of hypersurfaces. The additional requirements 
needed if the space-time is stationary but not static 
are that the black hole is axisymmetric and invariant
under simultanous reflection of time and and the angle
around the symmetry axis.}

Before proceeding, let us explain why it is desirable
to admit actions containing higher derivative terms.
The reason is that we would like to include so-called
effective actions which incorporate quantum effects. 
In quantum field theory the effective action  
is defined to be the generating functional of the
correlation functions. Since the classical action
generates the classical contribution to the correlation
functions (the leading part in an expansion in $\hbar$)
the effective action might be considered to be 
its quantum version. Unfortunately the exact effective action
is usually a rather formal and inaccessible object.
However, certain approximations can be computed, and
string theory provides a framework where 
quantum corrections to the gravitational action can
and have been computed.\footnote{We refer to \cite{Habil}
for more details.}
 As expected on general grounds,
quantum gravity manifests itself in the form of higher
derivative terms in the effective action, in particular
terms which contain higher powers of the Riemann tensor
and its contractions. We will discuss a particular class
of such terms in the next section. 

Let us next turn to the first law of black hole
mechanics, which states that for a stationary 
black hole an infinitesimal change of the mass $M$ is 
related to infinitesimal changes of the horizon area
$A$, of the angular momentum $J$ and of the electric
charge $Q$ by:
\be
\delta M = \frac{\kappa_S}{8 \pi} \delta A + \omega
\delta J + \phi \delta Q \;,
\ee
where $\omega$ is the angular velocity and $\phi$ 
the electrostatic potential. This should be compared
to the first law of thermodynamics (for a grand canonical
ensemble),
\be
\delta E = T \delta {\cal S} - p \delta V + \mu \delta N \;,
\ee
where $E$ is energy, $T$ is temperature, ${\cal S}$ is entropy,
$p$ is pressure, $V$ is volume $\mu$ is chemical  potential
and $N$ is particle number. Given the relation between 
surface gravity and temperature, this suggests to interprete
the area of the event horizon as the entropy of the black hole. 
This is surprising, since the entropy of normal thermodynamical
systems is an extensive property, i.e., proportional to the
volume rather than the surface area. 

Like the zeroth law the first law can be derived for 
general covariant actions, under the same assumptions as
for the zeroth law. Moreover, the mass,
entropy, angular momentum and charge of the black hole 
are defined as surface charges, which are
obtained by integrating a so-called Noether two-form over a 
closed spatial surface \cite{Wald}. The Noether two-form is constructed
out of Killing vector fields according to a certain algorithm.
For the special case of Einstein gravity this definition
reduces to the usual ones (i.e., the Komar or ADM constructions
of mass and angular momentum, and the proportionality of
entropy and horizon area).

Finally, let us turn to the second law of black hole
mechanics, the Hawking area law. In contrast to the zeroth and
first law, one does not assume the space-time to be 
stationary. Rather it can be time-dependent, and include
processes such as the formation and fusion of black holes,
as long as the time evolution is `asymptotically predictable'\footnote{We 
refer to \cite{Wald1} for a precise definition and more
details.}). 
The second law then states that 
the total area of event horizons is non-decreasing,
\be
\delta A \geq 0 \;,
\ee
which is obviously analogous to the second law of thermodynamics,
which states the same for the entropy,
\be
\delta {\cal S} \geq 0 \;.
\ee
This reinforces the identification of area and entropy suggested
by the first law. The second law has been derived using Einstein's
field equation together with conditions on the energy-momentum 
tensor of matter (plus assuming `predictability' of space-time).
So far there is no general proof for  the case of general 
covariant actions. However, examples have been studied, and
the integrated Noether form is a good candidate for entropy
in non-stationary space-times \cite{IyeWal:1994}.
One interesting question is whether one should expect that 
the second law holds for all covariant actions. Since dynamical
processes such as collision of black holes are admitted,
the contents of the second law appears to be more sensitive
to the details of the dynamics as the zeroth and first law.
It is not clear whether all possible higher derivative actions
give rise to `sensible' physics which respects the second law.
But one would certainly expect this to be true for string-effective
actions, though this does not seem to have studied so far. 
Anyway, already the zeroth and first law provide compelling
evidence for relating 
relating the surface gravity to the temperature and the area 
(integrated Noether two-form) to the entropy.

\subsection{Quantum aspects of black holes}

Let us now review the role of the Hawking effect \cite{Haw} in making 
plausible
the re-interpretation of geometrical as thermodynamic properites.
This effect is derived
by treating space-time geometry as a classical background,
while matter is described by quantum field theory. In this
framwork it has been shown that black holes emit thermal radiation,
even if there is no ingoing radiation or matter. Moreover, the
so-called Hawking temperature of this radiation is indeed
proportional to the surface gravity:
\be
T_{\rm Hawking} = \frac{\kappa_S}{2 \pi} \;.
\ee
In Einstein gravity the factor of proportionality between area and
entropy is then fixed by the first law:
\be
{\cal S} = \frac{A}{4} 
\ee
(Newton's and Planck's constant and the speed of light serve
as natural units, $G_N = \hbar = c = 1$).
When using a covariant higher derivative action, the
entropy is given by integrating the Noether two form $Q$ over the
horizon $H$:
\be
{\cal S} = 2 \pi \oint_H Q  \;.
\ee
It has been shown that the entropy can be expressed in terms of 
variational derivatives of the Lagrangian with respect to 
the Riemann tensor \cite{JacKanMye,IyeWal:1994}:
\be
{\cal S} = 2 \pi \oint_H \frac{\delta {\cal L}}{\delta R_{\mu \nu \rho 
\sigma}} \varepsilon^{\mu \nu} \varepsilon^{\rho \sigma} \,\sqrt{h} \,
d^2 \Omega \;,
\label{VarFor}
\ee
where $\epsilon^{\mu \nu}$ is the normal bivector of $H$ (with a 
certain normalization), and $\sqrt{h} \, d^2 \Omega$ is the induced
volume element. If ${\cal L}$ is 
the Einstein-Hilbert Lagrangian, this formula reproduces the area 
law. If further terms containing the Riemann tensor are present
in ${\cal L}$ they 
induce explicit modifications of the area law.

Once the Hawking effect is taken into account, black holes can
emit radiation, which implies that they loose mass and shrink,
thus violating the second law of black hole mechanics. However,
as soon as one takes the idea seriously that black holes carry
entropy, one should consider the total entropy obtained by 
adding black hole entropy and the thermodynamical entropy of 
the exterior region. The generalized
second law of thermodynamics, which states that the total entropy
is non-decreasing, is expected to be 
valid in quantum gravity \cite{Bekenstein}.

So far we have considered black hole entropy from what one might
call the macroscopic or thermodynamical perspective. When dealing
with many-constitutent systems one distinguishes two
levels of description. The fundamental or microscopic level
of description requires knowledge of the precise state of the 
system. For a classical gas this would require to
specify the positions and momenta (and other quantities if
internal excitations exist)
of all atoms or molecules. At the thermodynamical or macroscopic
level of description one only considers collective  
properties of the system, such as temperature, volume and pressure. 
Statistical mechanics asserts that these macroscopic quantities
arise by `coarse graining' microscopic quantities. E.g., 
temperature is the average energy per degree of freedom. Obviously
many microstates will give rise to the same macrostate, where the
latter is characterised by fixing only the macroscopic
quantities. The so-called statistical or microscopic entropy
measures how many different microstates give rise to the same
macrosate. If $d(E,\ldots)$ denotes the number of microstates 
corresponding to the macrostate with energy $E$, etc., then
the corresponding microscopic entropy is 
\be
{\cal S}_{\rm micro} = \log d(E, \ldots) \;.
\ee
In contrast the so-called macroscopic or thermodynamical entropy
${\cal S}_{\rm macro}$ 
is a purely macroscopic quantity, which can be characterized by
its relation to other macroscopic quantities, such as temperature,
free energy, etc. Both entropies are expected to be equal in the
thermodynamcial limit, i.e., when the number of constituents goes
to infinity. 

The geometrical black hole entropy is analogous to the 
macroscopic entropy, because it has been defined through 
relations which only involve collective properties of the black hole,
such as mass, charge and angular momentum. Any
theory of quantum gravity is expected to provide a corresponding
microscopic description of black holes, which in particular 
allows one to identify its microstates. In particular 
the microscopic entropy should be equal to the macroscopic one,
at least in the limit of large mass, which is analogous to the
thermodynamical limit. This is widely regarded as a benchmark test
for theories of quantum gravity. 

\subsection{Black holes and strings}

In string theory four-dimensional black holes can be interpreted 
as arising from states in the full ten-dimensional string
theory. These states might be string states, 
or winding states of higher-dimensional membranes 
(i.p. D-branes) \cite{BHcorr}. 
One can
test the expected relation between macroscopic and microscopic entropy
by counting the ten-dimensional
states which give rise to the same four-dimensional black hole.
This comparison generically involves the variation of
parameters such as the string coupling, and it is not a priori clear
whether the number of states is preserved under this interpolation.
But for a special subclass of states, the so-called supersymmetric
states or BPS states, which we will review below,
the interpolation is at least highly plausible.
Moreover, both the macroscopic and the microscopic entropy can often be
computed to high precision and it has been found that they
match \cite{StrVaf}, even when subleading corrections are 
included \cite{CdWM:98}. 
In particular
these tests are sensitive to the distinction between the area law
and the generalized formula (\ref{VarFor}), and clearly show that string theory
`knows' about the modifications of the area law. In performing 
these prescision tests, special geometry plays a central role. 
It is the indispensible tool for constructing black hole solutions
and extracting the macroscopic entropy from them. This will be the
subject of the next section. We will not be able to cover the
microscopic side of the story, i.e., the counting of microstates,
in this article. 

\subsection{Black holes and supersymmetry}

Before turning to the details, let us review the concepts of 
BPS states and BPS solitons.\footnote{In the following
we use basic facts about supersymmetry algebras and their
representations. See for example \cite{WaB}, Chapter II.}
Recall that the supercharges 
which generate supersymmetry transformations are spinors. 
If there is more than one such spinor, than the supersymmetry
algebra admits central operators, which can be organised 
into a complex antisymmetric matrix. The skew eigenvalues 
$Z_{(i)}$ of this matrix are called the central charges.
It can be shown that on any irreducible representation the
mass is bounded from below by the absolute values of the 
charges:
\be
M \geq |Z_{(1)}| \geq |Z_{(2)}| \geq \cdots \;.
\ee 
Moreover, when the mass saturates one or several of these bounds,
part of the supercharges operate trivially, and the corresponding
multiplet is shorter than a generic massive multiplet. Such multiplets
are called supersymmetric multiplets or BPS multiplets.
When all inequalities are saturated, the resulting
BPS multiplet is invariant under half of the supertransformations
and  has as many states as a massless multiplet.  In the case of
$N=2$ supersymmetry considered in this article, the algebra has
one single complex supercharge $Z$. Consequently, there are 
generic massive supermultiplets $M > |Z|$ and `$\ft12$-BPS multiplets'
with $M=|Z|$. 

The concept of BPS state can be applied to solitons. By solitons
we refer to  solutions of the field equations which can be interpreted
as particle-like objects. In particular, these solutions are 
required to have finite energy, and therefore must approach the 
ground state asymptotically. Since the energy 
localized in a small part of space-time, such `lumps'
can be thought of as `extended particles'. One also requires 
that the solution is static (describing `a massive particle in its
rest frame') and free of naked singularities (we admit singularities
covered by event horizons in order to include black holes). 
A soliton is then called supersymmetric or BPS, if it is invariant
under part of the supersymmetry transformations. Let us denote
the fields of the underlying action collectively by $\Phi$, 
the spinorial supersymmetry transformation parameters by $\epsilon$,
the corresponding supersymmetry transformation by $\delta_\epsilon$
and the soliton solution by $\Phi_0$. Then a solution is BPS
if there exists a choice of $\epsilon$ such that
\be
(\delta_\epsilon \Phi)_{|\Phi_0} = 0  \;.
\ee
Particular examples of BPS solitons are provided by
supersymmetric black hole solutions of supergravity actions.
In supergravity the supersymmetry transformation parameters 
depend on space-time, $\epsilon=\epsilon(x)$. Therefore the BPS
condition implies the existence of a spinor field which generates
a supertransformation under which the black hole solution is invariant.
This is analogous to a Killing vector field, which generates a diffeomorphism
under which the metric (and possibly other fields) are invariant. 
Therefore such spinor fields $\epsilon(x)$ are called Killing spinors
(more accurately Killing spinor fields). The interested reader
is referred to the monograph \cite{Ortin} for a detailed discussion
of supersymmetric solutions.

\section{Special geometry and black holes}

\subsection{Vector multiplets coupled to gravity}

We are now in position to discuss  
BPS black hole solutions in $N=2$ supergravity coupled
to $n$ vector multiplets. This is the relevant part of 
the effective action for string compactifications preserving
$N=2$ supersymmetry. The general $N=2$ vector multiplet action
was constructed using the
superconformal calculus \cite{deWLauVanP}.\footnote{Further
references on $N=2$ vector multiplet Lagrangians and the
superconformal calculus include \cite{deWvHvP,deWLauvP,deWvP,CKvPDFdeWG}.} 
The idea of this method
is to start with a theory of $n+1$ rigidly supersymmetric
vector multiplets and to impose that the theory 
is invariant under superconformal transformations. 
This implies that the prepotential has to be homogenous
of degree 2 in addition to being holomorphic:
\be
F(\lambda X^I) = \lambda^2 F(X^I) \;,\;\;\;\lambda \in 
\mathbbm{C}^* \;,
\ee
where now $I=0,1, \ldots, n$. Next one `gauges' the
superconformal transformation, that is one makes
the Lagrangian locally superconformally invariant by
introducing suitable connections. The new fields
entering through this process are encoded in the
so-called Weyl multiplet.\footnote{One also needs
to add a further `compensating multiplet', which
can be taken to be a hypermultiplet. We won't need
to discuss this technical detail here. See for example
\cite{Habil} for more background material and references.} 
Finally, one imposes
gauge conditions which reduce the local superconformal
invariance to a local invariance under standard
(Poincar\'e) supersymmetry. Through the gauge
conditions some of the fields become functions of 
the others.
In particular, only $n$ out of the $n+1$ complex
scalars are independent. A convenient
choice for the independent scalars is
\be
z^A = \frac{X^A}{X^0} \;,
\label{SpecialCoord}
\ee
where $A=1,\ldots, n$.
This provides a set of special coordinates for the
scalar manifold $M_{VM}$. In contrast, all $n+1$
gauge fields remain independent. While one particular
linear combination, the so-called graviphoton, belongs
to the Poincar\'e supergravity multiplet, the
other $n$ gauge fields sit in vector multiplets,
together with the scalars $z^A$. 
The Weyl multiplet also provides  
physical degrees of freedom, namely the graviton and
two gravitini. 

From the underlying rigidly superconformal
theory the supergravity theory inherits the
invariance under symplectic rotations. For the
gauge fields this is manifest, as $(F^I_{mn}, G_{I|mn})$
transforms as a vector under $Sp(2(n+1), \mathbbm{R})$.\footnote{
The dual gauge fields $G_{I|mn}$ were introduced at
the beginning of section 2.}
In the scalar sector $(X^I, F_I)$, where
$F_I = \der_I F$, also transforms as a vector, while
the gravitational degrees of freedom are invariant.
To maintain manifest symplectic invariance, it is
advantagous to work with  $(X^I, F_I)$ instead
of $z^A$. 

The underlying geometry can be described as follows
\cite{Freed,Hitchin,ACD}: the fields $X^I$ provide coordinates
on the scalar manifold of the associated rigidly 
superconformal theory. This manifold has complex
dimension $n+1$, and can be immersed into
$T^* \mathbbm{C}^{n+1} \simeq \mathbbm{C}^{2(n+1)}$
just as described in the previous section. The additional
feature imposed by insisting on superconformal invariance
is that the prepotential is homogenous of degree 2. 
Geometrically this implies that the resulting affine special
K\"ahler manifold is a complex cone. The scalar manifold
of the supergravity theory is parametrized by the
scalars $z^A$ and has complex dimension $n$. It is
obtained from the manifold of the rigidly superconformal
theory by gauge-fixing the dilatation and $U(1)$
symmetry contained in the superconformal algebra.
This amounts to taking the quotient of the complex cone
with respect to the $\mathbbm{C}^*$-action $X^I \rightarrow
\lambda X^I$. Thus the
scalar manifold $M_{VM}$ is the basis of the 
conical affine special K\"ahler manifold $C(M_{VM})$
of the rigid theory. For many purposes, including
the study of black hole solutions, it is advantagous
to work on $C(M_{VM})$ instead of $M_{VM}$. In particular,
this allows to maintain manifest symplectic covariance,
as we already noted. In physical terms this means that 
one can postpone the gauge-fixing of the dilatation and $U(1)$
transformations. The manifolds which can be obtained
from conical affine special K\"ahler manifolds by 
a $\mathbbm{C}^*$-quotient are called projective
special K\"ahler manifolds. These are the target spaces
of vector multiplets coupled to supergravity. All 
couplings in the Lagrangian and all relevant geometrical data of
$M_{VM}$ are encoded in the prepotential. In particular, 
the affine special K\"ahler metric on $C(M_{VM})$ has
K\"ahler potential 
\be
K_C(X^I, \overline{X}^I) = 
-i ( X^I \overline{F}_I - F_I \overline{X}^I) \;,
\label{KCXXb}
\ee
while the projective special K\"ahler metric on $M_{VM}$ has
K\"ahler potential
\be
K(z^A, \overline{z}^{\overline{B}}) = - \log 
\left( -i ( X^I \overline{F}_I - F_I \overline{X}^I) \right)\;,
\label{KXXbar}
\ee
with corresponding metric 
\be
g_{a \overline{b}} = \frac{ \der^2 K(z^A, \overline{z}^{\overline{B}} )}
{\der z^a \der \overline{z}^b } \;.
\ee

In string theory the four-dimensional supergravity Lagrangians
considered here are obtained by dimensional reduction of the
ten-dimensional string theory on a compact six-dimensional
manifold $X$ and restriction to the massless modes. Then the
scalar manifold $M_{VM}$ is the moduli space of $X$. 
It turns out that the moduli spaces of Calabi-Yau threefolds
provide natural realizations of special K\"ahler geometry 
\cite{Strominger}. Consider for instance the Calabi-Yau
compactification of 
type-IIB string theory. In this
case $M_{VM}$ is the moduli space of complex structures of $X$,
the cone $M_{VM}$ is the moduli space of complex structures 
together with a choice of the holomorphic top-form, and
$T^* \mathbbm{C}^{n+1} \simeq \mathbbm{C}^{2(n+1)}$
is $H^3(X,\mathbbm{C})$, see \cite{Cor1}.

\subsection{BPS black holes and the attractor mechanism}

Let us then discuss BPS black hole solutions of
$N=2$ supergravity with $n$ vector multiplets. 
These are static, spherically symmetric
solutions of the field equations, which are asymptotically
flat, have regular event horizons, and possess 4 Killing
spinors. Since the $N=2$ superalgebra has 8 real supercharges,
these are $\ft12$-BPS solutions.
Let us first have a look at pure four-dimensional 
$N=2$ supergravity, i.e., we drop the vector multiplets, $n=0$.
The bosonic part of this theory is precisely the 
Einstein-Maxwell theory. In pure $N=2$ supergravity,
BPS solutions have been classified
\cite{GibHul,Tod,KowGlik}. The number of linearly independent
Killing spinor fields can be 8,4 or 0. This can be seen,
for example, by investigating the integrability conditions of the
Killing spinor equation.\footnote{The classification of supersymmetric
solutions has recently moved to the focus of interest. Readers
who want to get an idea how the classification of supersymmetric 
solutions of four-dimensional $N=2$ supergravity
would work with `modern', systematic
methods can consult \cite{Gauntlett:2002nw}, where all supersymmetric
solutions of minimal five-dimensional  supergravity were constructed.}
Solutions with
8 Killing spinors are maximally supersymmetric and
therefore considered as supersymmetric ground states.
Examples are Minkowski space and $AdS^2 \times S^2$. Solutions 
with 4 Killing spinors are called $\ft12$-BPS, because
they are invariant under half as many supersymmetries as
the ground state. They are solitonic realisations of
states sitting in BPS representations. 
For static $\ft12$-BPS solutions 
the space-time metric takes the form \cite{GibHul,Tod}
\be
ds^2 = - e^{-2f(\vec{x})} dt^2 + e^{2f(\vec{x})} d\vec{x}^2 \;,
\label{Isotropic}
\ee
where $\vec{x}=(x_1,x_2,x_3)$ are space-like coordinates
and the function $f(\vec{x})$ must be such that
$e^{f(\vec{x})}$ is a harmonic function with respect to $\vec{x}$.
The solutions also have a non-trivial gauge field, which likewise can
be expressed in terms of $e^{f(\vec{x})}$. This class
of solutions of Einstein-Maxwell theory is known as the 
Majumdar-Papapetrou solutions \cite{Maj,Pap}. The only Majumdar-Papapetrou 
solutions without naked singularities are the multi-centered
extremal Reissner-Nordstrom solutions, which describe static
configurations of extremal black holes, see for example \cite{Chandra}.
If one imposes in addition spherical symmetry, one arrives
at the extremal Reissner-Nordstrom solution describing a single
charged black hole. In this case the metric takes the 
form 
\be
ds^2 = - e^{-2f(r)} dt^2 + e^{2 f(r)} (dr^2 + r^2 d \Omega^2) \;,
\ee
where $r$ is a radial coordinate and  $d\Omega^2$ 
is the line element on the unit two-sphere.
The harmonic function takes the form
\be
e^{f(r)} = 1 + \frac{q^2 + p^2}{r} \;,
\label{isotropic}
\ee
where $q,p$ are the electric and magnetic charge with respect to 
the graviphoton.
The solution has two asymptotic regimes. In one limit,
$r\rightarrow \infty$, it becomes asymptotically flat:
$e^f \rightarrow 1$. In the other limit, $r\rightarrow 0$,
which is the near-horizon limit, it takes the
form
\be
ds^2 = - \frac{r^2}{q^2+p^2 } dt^2  +
\frac{q^2+p^2 }{r^2} dr^2 + (q^2 + p^2) d \Omega^2  \;.
\ee
This is a standard form for the metric of $AdS^2 \times S^2$.
The area of the two-sphere, which is the area of the event horizon 
of the black hole, is given by $A= 4 \pi (q^2 + p^2)$.
The two limiting solutions, flat Minkowski space-time
and $AdS^2 \times S^2$ are among the fully supersymmetric solutions
with 8 Killing spinors that we mentioned before. Thus,
the extremal Reissner-Nordstrom black hole interpolates
between two supersymmetric vacua \cite{Gibbons}. This is
a property familiar from two-dimensional kink solutions,
and motivates the interpretation of supersymmetric black 
hole solutions as solitons, i.e., as particle-like collective
excitations.

Let us  now return to $N=2$ supergravity with an arbitrary number
$n$ of vector fields. We are interested in solutions which
generalize the extremal Reissner-Nordstrom solution. Therefore
we impose that the solution should be $\ft12$-BPS, static,
spherically symmetric, asymptotically flat, and that it should
have a regular event horizon.\footnote{This excludes both 
naked singularties and null singularities, where the horizon 
coincides with the singularity and has vanishing area.} 
More general $\ft12$-BPS solutions have been studied extensively in
the literature, in particular in \cite{BLS} and \cite{CdWKM:00}.
Recently, the classification of all $\ft12$-BPS solutions 
was achieved in \cite{MesOrt}.

BPS black holes in theories with $n$ vector multiplets depend
on $n+1$ gauge fields and on $n$ 
scalar fields. For any $\ft12$-BPS
solution, which is static and spherically symmetric, the 
metric can be brought to the form (\ref{isotropic}) \cite{CdWKM:00}.
The condition that the solution is static
and spherically symmetric is understood in the strong sense,
i.e., it also applies to the gauge fields and scalars. 
Thus gauge fields and scalars are
functions of the radial coordinate $r$, only. Moreover
the electric and magnetic fields are spherically symmetric,
which implies that each field strength $F^I_{mn}(r)$ has only
two independent components (see for example Appendix A of
\cite{Habil} for more details). 

The electric and magnetic charges carried by the solution are
defined through flux integrals of the field strength over
asymptotic two-spheres:
\be
(p^I, q_I) = \frac{1}{4\pi} \left( \oint F^I, \oint G_I \right) \;,
\ee
where $F^I, G_I$ are the two-forms associated with
the field strength $F^I_{mn}$ and their duals
$G_{Imn}$. As a consequence, the charges transform as
a vector under symplectic transformations. By contracting
the charges with the scalars one obtains the symplectic
function
\be
Z = p^I F_I - q_I X^I \;.
\ee
This field is often called the central charge, which is
a bit misleading because $Z$ is a function of the fields
$X^I$ and $F_I$ and therefore a function of the scalar fields
$z^A$, which are space-time dependent.\footnote{One can
analyse BPS solutions without imposing the gauge conditions
which fix the superconformal symmetry, and in fact it is
advantagous to do so \cite{CdWM:98,CdWKM:00}. 
Then the scalars are encoded in the fields $X^I(r)$, which  
are subject to gauge transformations. 
Once gauge conditions are imposed, one can 
express $Z(r)$ in terms of the physical scalar fields $z^A(r)$. 
See \cite{Habil} for more details.} 
Hence, in the backgrounds
we consider, $Z$ is a function of the radial coordinate $r$.
However, when evaluating this field in the asymptotically
flat limit $r \rightarrow \infty$, it computes the electric
and magnetic charge carried by the graviphoton, which combine
into the complex central charge of the $N=2$ algebra \cite{Teitel}.

In particular, the mass of the black hole is given by
\be
M = |Z|_{\infty} = M(p^I, q_I, z^A(\infty)) \;.
\ee
Thus BPS black holes saturate the
mass bound implied by the supersymmetry algebra.
Note that the mass does not only depend on the
charges, but also on the values of the scalars at
infinity, which can be changed continuously. 

The other asymptotic regime is the event horizon.
If the horizon is regular, then the solution 
must be fully supersymmetric in this 
limit \cite{AttractorEqs}. Thus,
while the bulk solution has 4 Killing spinors,
both asymptotic limits have 8. 
In the near horizon limit, 
the metric (\ref{isotropic}) takes the form
\be
ds^2 = - \frac{r^2}{|Z|^2_{\mscr{hor}}} dt^2  +
\frac{|Z|^2_{\mscr{hor}}}{r^2} dr^2 + |Z|^2_{\mscr{hor}} d \Omega^2  \;,
\ee
where $|Z|^2_{\mscr{hor}}$ is the value of $|Z|^2$ at the horizon.
As in the extremal Reissner-Nordstrom solution, this is 
$AdS^2 \times S^2$.
The area of the two-sphere, which is the area of the event horizon, 
is given by $A = 4 \pi |Z|^2_{\mscr{hor}}$. Hence the Bekenstein Hawking
entropy is 
\be
\Smacro = \frac{A}{4} = \pi |Z|^2_{\mscr{hor}} \;.
\ee
A priori, $\Smacro$ depends on both the
charges and the values of the scalars at the horizon,
and one might expect that one can change the latter
continuously. This would be incompatible with relating 
$\Smacro$ to a statistical entropy $\Smicro$ which 
counts states. But it turns out that the values of the
scalar fields at the horizon are themselves determined
in terms of the charges. Here, it is convenient to define
$Y^I = \overline{Z} X^I$ and
$F_I = F_I(Y) =   \overline{Z} F_I(X)$.\footnote{
Note that $F_I$ is homogenous of degree 1.}
In terms of these variables, the black hole attractor
equations \cite{AttractorEqs}, which express the horizon
values of the scalar fields in terms of the charges, take
the following, symplectically covariant form:
\be
\left( \begin{array}{c} 
Y^I - \overline{Y}^I \\
F_I - \overline{F}_I \\
\end{array} \right)_{\mscr{hor}} = i 
\left( \begin{array}{c} 
p^I \\ q_I \\
\end{array} \right) \;.
\label{AttractorEqs}
\ee

The name attractor equations refers to the behaviour
of the scalar fields as functions of the space-time
radial coordinate $r$. While the scalars can take
arbitrary values at $r \rightarrow \infty$, they flow
to fixed points, which are determined by the charges,
for $r\rightarrow 0$. This fixed point behaviour
follows when imposing that the 
event horizon is regular. Alternatively, one can show that
to obtain a fully supersymmetric solution with
geometry $AdS^2 \times S^2$ the scalars need 
to take the specific values dictated by the
attractor equations \cite{CdWKM:00}. This is due to the
presence of non-vanishing gauge fields.
The gauge fields in $AdS^2 \times S^2$ are
covariantly constant, so that this can be viewed
as an example of a flux compactification.
In contrast, Minkowski space is also maximally supersymmetric,
but the scalars can take arbitrary constant values, because
the gauge fields vanish. In type-II Calabi-Yau compactifications,
the radial dependence of the scalar fields defines a flow on
the moduli space, which starts at an arbitrary point and 
terminates at a fixed point corresponding to 
an `attractor Calabi Yau.' Since the electric and magnetic
charges $(p^I, q_I)$, which determine the fixed point, take
discrete values, such attractor threefolds sit at very 
special points in the moduli space. This has been studied in detail 
in \cite{Moore}.

Using the fields $Y^I$ instead of $X^I$ to parametrize
the scalars simplifies formulae and 
has the advantage that the $Y^I$ are invariant
under the $U(1)$ transformations of the superconformal
algebra. Note that
\be 
|Z|^2 = p^I F_I - q_I Y^I \;,
\label{Zabssquared}
\ee
which is easily seen using the homogeneity properties of
the prepotential. The diffeomorphism 
$X^I \rightarrow Y^I$ acts non-holomorphically
on the cone $C(M_{VM})$, but operates
trivially on its basis $M_{VM}$. Note in particular
that 
\be
z^A = \frac{X^A}{X^0} = \frac{Y^A}{Y^0} \;.
\label{zAXIYI}
\ee

\subsection{The black hole variational principle}

We now turn to the black hole variational principle,
which was found in \cite{BCdWKLM} and generalized  
in \cite{CdWKM:06}, motivated by the
observations of \cite{OSV}. First, we define
two symplectic functions,  
the entropy function 
\be
\Sigma(Y^I, \overline{Y}^I, p^I, q_I) = 
{\cal F}(Y^I, \overline{Y}^I) - q_I (Y^I + \overline{Y}^I)
+ p^I (F_I + \overline{F}_I)
\label{Sigma}
\ee
and the black hole free energy
\be
{\cal F}(Y^I, \overline{Y}^I) = -i \left(
\overline{Y}^I F_I - Y^I \overline{F}_I \right) \;.
\ee
The reason for our choice of terminology will become
clear later. Now we impose that the entropy function
is stationary, $\delta \Sigma =0$, under variations of
the scalar fields $Y^I \rightarrow Y^I + \delta Y^I$.
Using that the prepotential is homogenous of degree two, 
it is easy to see that the conditions for $\Sigma$ being
stationary are precisely the black hole attractor
equations (\ref{AttractorEqs}). Furthermore, at the
attractor point we find that\footnote{
The relation $- i\left(
\overline{Y}^I F_I - Y^I \overline{F}_I \right)
= \left( q_I Y^I - p^I F_I \right)$ follows from the definitions
of $Z$ and $Y^I$ together with the homogenity of the prepotential
(once the dilatational symmetry of the fields $X^I$ has been gauge fixed).
Therefore it holds irrespective of whether the scalar fields take
their attractor values or not.}
\bea
{\cal F}_{\mscr{attr}} &=& - i\left(
\overline{Y}^I F_I - Y^I \overline{F}_I \right)_{\mscr{attr}}
= \left( q_I Y^I - p^I F_I \right)_{\mscr{attr}} \nonumber \\
&=& \left( q_I \overline{Y}^I - p^I \overline{F}_I \right)_{\mscr{attr}} =
 - | Z |^2_{\mscr{attr}}
\eea
and therefore
\be
\Sigma_{\mscr{attr}} = | Z |^2_{\mscr{attr}} = \ft{1}{\pi} 
\Smacro(p^I, q_I) \;.
\ee
Here and in the following we use the label `attr' (instead of `hor' used
previously) 
to indicate that quantities are evaluated at the attractor point
determined by the electric and magnetic charges.

Thus, up to a constant factor, the entropy is obtained by 
evaluating the entropy function at its critical point. Moreover, a
closer look at the variational principle shows us that, again up to
a factor, the black hole entropy $\Smacro(p^I,q_I)$ is 
the Legendre transform of the free energy 
${\cal F}(Y^I, \overline{Y}^I)$, where the latter is considered 
as a function of $x^I = {\rm Re}(Y^I)$ and $y_I = {\rm Re}(F_I)$.
At this point the real variables discussed in the previous
section become important again. Note that the change of 
variables $(Y^I, \overline{Y}^I) 
\rightarrow (x^I, y_I)$ is well defined provided that
${\rm Im} (F_{IJ})$ is non-degenerate. This assumption
will be satisfied in general, but breaks down in
certain  string theory applications,  where one reaches 
the boundary of the moduli space.\footnote{See for example \cite{CdWKM:06} 
for a discussion of some of the implications.} 

We are therefore led to rewrite the variational principle
in terms of real variables. First, recall that the
Hesse potential $H(x^I, y_I)$ is the Legendre transform
of (two times) the imaginary part of the prepotential, see
(\ref{HessePrepot}).\footnote{Note that this is the 
Hesse potential of the affine special K\"ahler metric 
on $C(M_{VM})$. The projective special K\"ahler metric
on $M_{VM}$ is obtained by the $\mathbbm{C}^*$-quotient.} 
This Legendre transform 
replaces the independent variables $(x^I,u^I)$$ =$
(${\rm Re}(Y^I)$, ${\rm Im}(Y^I)$) by the independent
variables $(x^I,y_I)$$ =$( ${\rm Re}(Y^I)$, ${\rm Re}(F_I)$)
and therefore implements the change of variables
$(Y^I, \overline{Y}^I)  \rightarrow (x^I, y_I)$.
Using (\ref{HessePrepot}) we find 
\be
H(x^I, y_I) = - \ft{i}{2} (\overline{Y}^I F_I 
- \overline{F}_I Y^I ) = \ft12 {\cal F}(Y^I, \overline{Y}^I) \;.
\ee
Thus, up to a factor, the Hesse potential is the 
black hole free energy. 
We can now express the entropy function 
in terms of the real variables:
\be
\Sigma(x^I, y_I, p^I, q_I) = 2 H(x^I, y_I) - 2 q_I x^I 
+ 2 p^I y_I \;.
\label{SigmaReal}
\ee
If we impose that $\Sigma$ is stationary with respect to
variations of $x^I$ and $y_I$, we get the black hole
attractor equations in real variables:
\be
\frac{\der H}{\der x^I} = q_I \;,\;\;\;
\frac{\der H}{\der y_I} = - p^I \;.
\label{AttrEqsReal}
\ee
Plugging this back into the entropy function we obtain
\be
\Smacro = 2 \pi \left( H - x^I \frac{\der H}{\der x^I} 
- y_I \frac{\der H}{\der y_I} \right)_{\mscr{attr}} \;.
\label{SmacroReal}
\ee
Thus, up to a factor, the black hole entropy is the
Legendre transform of the Hesse potential. 
This is an intriguing observation, because it relates
the black hole entropy, which is a space-time quantity,
directly 
to the special geometry encoding the scalar dynamics.
In string theory compactifications this relates the
geometry of four-dimensional space-time to the geometry
of the compact internal space $X$. 
The Hesse potential
appears to be closely related to the action functional 
underlying the geometry of $X$ in Hitchin's approach to 
manifolds with special holonomy \cite{Hit2,PesWit,Pes}.

We can also relate the black hole free energy to another
quantity of special geometry. In terms of complex variables
we observe that
\be
{\cal F}(Y^I, \overline{Y}^I) =
K_C(Y^I, \overline{Y}^I) := 
i(\overline{Y}^I F_I - \overline{F}_I Y^I)  \;.
\ee
Comparing to (\ref{KCXXb}) it appears that we 
should interpet $K_C(Y^I, \overline{Y}^I)$ as the
K\"ahler potential of an affine special K\"ahler metric
on $C(M_{VM})$. Since the diffeomorphism $X^I \rightarrow
Y^I$ is non-holomorphic, this is not the same special 
K\"ahler structure as with  (\ref{KCXXb}). However, we 
already noted that the diffeomorphism acts trivially 
on $M_{VM}$, see (\ref{zAXIYI}). Moreover it is easy to see
that when taking the quotient with respect to the 
$\mathbbm{C}^*$-action $Y^I \rightarrow \lambda Y^I$, then 
the resulting projective special K\"ahler metric 
with K\"ahler potential $K(Y^I, \overline{Y}^I) = - \log
K_C(Y^I, \overline{Y}^I)$ is the
same as the one derived from (\ref{KXXbar}), because the
two K\"ahler potentials differ only by a K\"ahler transformation.
It appears that in the context of black hole solutions the
affine special K\"ahler metric associated with the rescaled
scalars $Y^I$ is of more direct importance than the one
based on the $X^I$. The same remark applies to the 
Hesse potential, which depends on the real coordinates associated
to $Y^I$. Note that the scalars $Y^I$ do not only encode
the values of the Calabi-Yau moduli $z^A$ via (\ref{zAXIYI})
but also, via  (\ref{Zabssquared})
the size of the two-sphere in the black-hole
space-time.\footnote{This is not only true at the horizon but
throughout the whole black hole solution.} While variations
of the moduli correspond to variations along the basis
of the cone $C(M)$, variations of the radius of the 
two-sphere correspond to motions along the radial direction
of the cone.

Note that it is more natural to identify the free
energy with the Hesse potential than the K\"ahler potential.
The first reason is that the various Legendre transforms
involve the real and not the complex coordinates. The second
reason is that, as we will discuss below, we need to generalize
the supergravity Lagrangian in order to take into account
certain corrections appearing in string theory. We will see
that this works naturally by introducing a generalized Hesse
potential. 

Before turning to this subject, we also remark that
the terms in the entropy function (\ref{Sigma}) which are linear
in the charges, and which induce the Legendre transform,
have yet another interpretation in terms of supersymmetric
field theory. Namely, the symplectic function 
\be
W = q_I Y^I - p^I F_I
\ee
has the form of an $N=2$  superpotential. The  four-dimensional
supergravity Lagrangian we are studying does not have
a superpotential. However, the near-horizon solution
has the form $AdS^2 \times S^2$ and carries non-vanishing,
covariantly constant gauge fields. The dimensional
reduction on $S^2$ is a flux compactification, with 
fluxes parametrized by $(p^I, q_I)$, and the resulting 
two-dimensional theory will possess a superpotential. 
This also provides an alternative interpretation of the
attractor mechanism, as the resulting scalar potential
will lift the degeneracy of the moduli.

\subsection{Quantum corrections to black holes solutions and
entropy}

So far we only considered supergravity Lagrangians which 
contain terms with at most two derivatives. The effective
Lagrangians derived from string theory also contain higher
derivative terms, which modify the dynamics at short distances.
These terms describe interactions between the massless states which
are mediated by massive string states. While the effective
Lagrangian does not contain the massive string states explicitly, it
is still possible to describe their impact on the dynamics of the
massless states. 

In $N=2$ supergravity a particular class of higher derivative
terms can be taken into account by giving the prepotential an
explicit dependence on an additional complex variable 
$\Upsilon$, which is proportional to the lowest component of the
Weyl multiplet \cite{AGNT,deW}. The resulting function $F(Y^I,\Upsilon)$ is 
required to be holomorphic in all its variables, and to be 
(graded) homogenous of degree two:\footnote{Since we are 
interested in black hole solutions, we use
rescaled fields $Y^I, \Upsilon$.}
\be
F(\lambda Y^I, \lambda^2 \Upsilon) = \lambda^2 F(Y^I, \Upsilon) \;.
\ee
Assuming that it is analytic at $\Upsilon =0$ one can expand it as
\be
F(Y^I, \Upsilon) = \sum_{g=0}^\infty F^{(g)}(Y^I) \Upsilon^g \;.
\label{Expand}
\ee
Then $F^{(0)}(Y^I)$ is the prepotential, while the functions 
$F^{(g)}(Y^I)$ with $g>0$ appear in the Lagrangian as the
coefficients of various higher-derivative terms. These include
in particular terms quadratic in the space-time curvature,
and therefore one often loosely refers to the higher derivative
tems as $R^2$-terms.

In type-II Calabi Yau compactifications the functions $F^{(g)}(Y^I)$
can be computed using (one of) the topologically twisted version(s) of the
theory \cite{BCOV}. 
They are related to the partition functions $Z^{(g)}_{\mscr{top}}$
of the topologically twisted string on a world sheet with genus
$g$ by $F^{(g)} = \log Z^{(g)}_{\mscr{top}}$. Therefore they are
called the (genus-$g$) topological free energies. 

It was shown in \cite{CdWM:98,CdWKM:00} that the black hole
attractor mechanism can be generalized to the case of 
Lagrangians based on a general function $F(Y^I,\Upsilon)$.
The attractor equations still take the form 
(\ref{AttractorEqs}), but the prepotential is replaced by 
the full function $F(Y^I,\Upsilon)$. The additional variable
$\Upsilon$ takes the value $\Upsilon=-64$ at the horizon.
The evaluation of the generalized entropy formula (\ref{VarFor}) for
$N=2$ supergravity gives \cite{CdWM:98}:
\be
\Smacro(q^I, p_I) = \pi \left( 
|Z|^2 + 4 \mbox{Im}(\Upsilon F_{\Upsilon}) \right)_{\mscr{attr}} \;,
\ee
where $F_\Upsilon = \der_\Upsilon F$.\footnote{At the attractor
point, $\Upsilon$ takes the value $\Upsilon=-64$.} Note that
symplectic covariance is manifest, as the entropy is the sum
of two symplectic functions. While the first term
corresponds to the area law, the second term is an explicit
modification which depends on the coefficients $F^{(g)}$, $g>0$, of
the higher derivative terms.

It was shown in \cite{CdWKM:06} that the variational
principle generalizes to the case with $R^2$-terms. The black hole free
energy ${\cal F}$ is now proportional to a generalized 
Hesse potential $H(x^I, y_I, \Upsilon, \overline{\Upsilon})$, which in turn 
is proportional to the Legendre transform of the imaginary part 
of the function $F(Y^I, \Upsilon)$:
\[
H(x^I, y_I, \Upsilon, \overline{\Upsilon}) = 
2\mbox{Im} F(x^I + i u^I , \Upsilon) - 2 y_I u^I  \;.
\]
In terms of complex fields $Y^I$ this becomes
\bea
H(x^I, y_I, \Upsilon, \overline{\Upsilon}) &=& 
- \ft{i}{2} (\overline{Y}^I F_I - \overline{F}_I Y^I)
-i ( \Upsilon F_\Upsilon - \overline{\Upsilon} 
\overline{F}_{\overline{\Upsilon}} ) \label{GenHesse} \\ 
&=& \ft12 {\cal F}(Y^I, \overline{Y}^I,
\Upsilon, \overline{\Upsilon}) \;. \nonumber
\eea
The entropy function (\ref{SigmaReal}), 
the attractor equations (\ref{AttrEqsReal}) and the formula for the 
entropy (\ref{SmacroReal}),
which now includes correction terms to the area law,
remain the same, except that one uses the generalized
Hesse potential. From (\ref{GenHesse}) it is obvious that
the black hole free energy naturally corresponds to a
generalized Hesse potential (defined by the Legendre
transform of the prepotential) and not to a 
`generalized K\"ahler potential', which would only give
rise to the first term on the right hand side of
(\ref{GenHesse}).

There is a second class of correction terms in string-effective
supergravity Lagrangians. Quantum corrections involving the
massless fields lead to modifications which correspond to 
adding non-holomorphic terms to the function $F(Y^I, \Upsilon)$.
The necessity of such non-holomorphic terms can be seen
by observing that otherwise the invariance of the full string theory 
under T-duality and S-duality is not captured by the effective
field theory. In particular, one can show 
that the black hole entropy can only be T- and  S-duality invariant 
if non-holomorphic corrections are taken into account 
\cite{CdWM:9906}.\footnote{We
are referring to compactifications with exact T- and S-duality
symmetry. These are mostly compactifications with $N=4$
supersymmetry, which, however, can be studied in the $N=2$
framework. We refer to \cite{CdWM:9906,CdWKM:04,CdWKM:06} for details.} 
From the point of view of string theory the presence of these
terms is related to a holomorphic anomaly \cite{BCOV,AGNT}.

As the holomorphic $R^2$-corrections, the non-holomorphic 
corrections can be incorporated into the black hole 
attractor equations and the black hole variational principle
\cite{CdWM:9906,CdWKM:06}. The non-holomorphic terms 
are encoded in a function $\Omega(Y^I, \overline{Y}^I,
\Upsilon, \overline{\Upsilon})$, which is real valued
and homogenous of degree two. To incorporate non-holomorphic
terms into
the variational principle one has to define the generalized
Hesse potential as the Legendre transform of 
$2 {\rm Im}F + 2 \Omega$:
\be
H(x^I, \hat{y}_I, \Upsilon, \overline{\Upsilon}) 
= 
2 {\rm Im} F(x^I + i u^I, \Upsilon, \overline{\Upsilon}) 
+ 2 \Omega (x^I, u^I, \Upsilon, 
\overline{\Upsilon}) - 2 \hat{y}_I u^I \;,
\label{nonharmonic}
\ee
where  $\hat{y}_I = y_I + i( \Omega_I - \Omega_{\overline{I}})$
and $\Omega_I = \ft{\der \Omega}{\der Y^I}$ and 
$\Omega_{\overline{I}} = \ft{\der \Omega}{\der \overline{Y}^I}$.
Up to these modifications, the attractor equations, the entropy
function, and the entropy remain as in (\ref{AttrEqsReal}),
(\ref{SigmaReal}) and (\ref{SmacroReal}). 
Also note from (\ref{nonharmonic}) that 
if $\Omega$ is harmonic, it can be absorbed into $\mbox{Im} F$, 
because it then is the imaginary part of holomorphic function.
Thus, the  non-holomorphic modifications of the prepotentail 
correspond to non-harmonic functions $\Omega$.

In terms of the complex variables the attractor equation are
\be
\left( \begin{array}{c}
Y^I - \overline{Y}^I \\
F_I + 2 i \Omega_I - F_I + 2i \Omega_{\overline{I}} \\
\end{array}\right) 
= i \left( \begin{array}{c} 
p^I \\ q_I \\ \end{array} \right) \;.
\label{AttrEqsNH}
\ee
The modified expressions for the free energy
and the entropy function can be found in \cite{CdWKM:06}.

At this point it is not quite clear what the 
$R^2$-corrections and the non-holomorphic corrections mean
in terms of special geometry. Since they correspond to higher
derivative terms in the Lagrangian, they do not give rise
to modifications of the metric on the scalar manifold, which,
by definition, is the coefficient of the scalar two-derivative
term.\footnote{See however \cite{ShiYin}, where such an 
interpretation was proposed.} It would be very interesting
to extend the framework of special geometry such that 
the functions $F^{(g)}$ get an intrinsic geometrical meaning.

\subsection{Black hole partition functions and the topological
string}

Let us now discuss how the black hole variational principle 
is related to black hole partition functions and the topological
string. We start by relating the variational principle described 
in the last sections to the variational principle used in 
\cite{OSV}. One can
start from the generalized Hesse potential and perform 
partial Legendre transforms by imposing only part of the
attractor equations. If this subset of fields is properly chosen 
one obtains a reduced variational principle, which yields 
the remaining attractor equations, and, by 
further extremisation, the black hole entropy.
Specifically, one can solve the magnetic attractor
equations $Y^I - \overline{Y}^I = i p^I$
by setting\footnote{Obviously, 
$\phi^I = 2 x^I$. We use $\phi^I$ to be consistent 
with the notation used in \cite{CdWKM:06}. The conventions
of \cite{OSV} are slightly different.}
\be 
Y^I = \ft12 (\phi^I + i p^I) \;.
\ee
Plugging this back, the new, reduced entropy function is
\be
\Sigma(p^I, \phi^I, q_I) = {\cal F}_E (p^I, \phi^I, \Upsilon, 
\overline{\Upsilon})  - q_I \phi^I \;,
\ee
where\footnote{We suppressed the dependence of $\Sigma$ on 
$\Upsilon$, but indicated it for ${\cal F}_E$ in order to 
make explicit that we included the higher derivative corrections.}
\be
{\cal F}_E (p^I, \phi^I, \Upsilon, \overline{\Upsilon} ) 
= 4 \left( \mbox{Im} F(Y^I, \Upsilon) 
+ \Omega(Y^I, \overline{Y}^I, \Upsilon, \overline{\Upsilon}) 
\right)_{\mscr{mgn}}
\ee
Here the label `mgn' indicates that the magnetic attractor 
equations have been imposed, i.e., $Y^I = \ft12 (\phi^I + i p^I)$. 
Both  ${\cal F}(Y^I, \overline{Y}^I, \Upsilon, \overline{\Upsilon}) = 
2 H(x^I, \hat{y}_I, \Upsilon, \overline{\Upsilon})$
and  ${\cal F}_E (p^I, \phi^I, \Upsilon, \overline{\Upsilon})$ 
are 
interpreted as free energies, which, however, refer to different
statistical ensembles. In the microcanonical ensemble the electric
and magnetic charges are kept fixed, while they fluctuate 
around a mean value in the canoncial ensemble. The transition
between these two ensemble is made by changing the independent
variables, i.e., one eliminates the electric and magnetic
charges $q_I, p^I$ in favour of the corresponding chemical 
potentials, which are the electrostatic
and magnetostatic potentials $\phi^I, \chi_I$.\footnote{Since 
the charges play the roles of
particle number in non-relativistic thermodynamics, it might
appear more logical to call the `microcanonical' ensemble 
canonical, and the `canonical' ensemble grand canonical. However,
we follow the terminology established in the recent literature
on the OSV conjecture.} By virtue of the equations of motion
the potentials coincide, up to a factor, with the real
coordinates on $C(M_{VM})$: $\phi^I = 2 x^I$, $\chi_I = 2 \hat{y}_I$.
In black hole thermodynamics the electrostatic and magnetostatic
potentials are evaluated at the horizon. Note that  both sets
of thermodynamical variables correspond to different real
symplectic coordinates on $C(M_{VM})$: the charges to the
imaginary part, the potentials to the real part of the
symplectic vector $(Y^I, F_I)$.

As an intermediate step, one can go to the mixed ensemble,
where the magnetic charges are kept fixed, while the electric
charges fluctuate. Then the independent variables are $p^I$ and
$\phi^I$. This indicates that ${\cal F}$ is the free energy
with respect to the canonical ensemble, while ${\cal F}_E$
is the free energy with respect to the mixed ensemble.

If one imposes that 
$\Sigma(p^I, \phi^I, q_I) $ is stationary with respect 
to variations of $\phi^I$, then one obtains the 
electric attractor equations 
$(F_I  - 2i \Omega_I) - 
(\overline{F}_I + 2i \Omega_{\overline{I}}) 
= i q_I$ (\ref{AttrEqsNH}).
Plugging these back one sees that at the
stationary point $\Sigma_{\mscr{attr}}
= \ft{1}{\pi} \Smacro(p^I, q_I)$ and that the
macroscopic entropy is the partial Legendre transform
of the free energy ${\cal F}_E(p^I, \phi^I, \Upsilon, \overline{\Upsilon})$.

Actually, the black hole free energy introduced in 
\cite{OSV} includes the contribution from holomorphic higher derivative 
terms, but not the non-holomorphic corrections. Let us
denote this quantity 
by ${\cal F}_{\rm OSV}(p^I,
\phi^I, \Upsilon)$. It is proportional to the
imaginary part of the generalized holomorphic prepotential
$F(Y^I, \Upsilon)$. If the model under consideration
has been obtained by compactification of type-II string theory
on a Calabi-Yau threefold, then 
the prepotential is in turn proportional to the 
so-called topological free energy $F_{\rm top}$, which 
is the logarithm of the all-genus
partition function of the topological type-II string,
$Z_{\rm top} = e^{F_{\rm top}}$. In our conventions 
the precise relation between the free energies is
\be
\pi {\cal F}_{\rm OSV} = 4 \pi \mbox{Im} F = 2 \mbox{Re} F_{\rm top} \;.
\ee
Therefore the free energy ${\cal F}_{\rm OSV}$ is related to 
the topological partition function by \cite{OSV}
\be
e^{\pi {\cal F}_{\rm OSV}(p, \phi, \Upsilon)}
= |Z_{\mscr{top}}|^2 \;. 
\label{Ztop} 
\ee
This supports the idea to take the interpretation of 
${\cal F}_{\rm OSV}(p,\phi, \Upsilon)$ as the free energy of the black hole 
seriously. Then it should be related to the partition function of
the black hole with respect to the mixed ensemble, which is 
defined by 
\be
Z_{\rm mixed}(p,\phi) = \sum_q d(p,q) e^{q \phi}  \;,
\ee 
where $d(p,q)$ is the number of BPS microstates with charges 
$p^I, q_J$, and  $q \phi := q_I \phi^I$. This relation is a 
formal discrete Laplace transform which relates the 
microscopic partition function, i.e., the state degeneracy, to
the mixed partition function.
The standard relation
between free energy and partition function would imply that
$Z_{\rm mixed} = e^{\pi {\cal F}_{\rm OSV}}$. However, from our 
discussion of the black hole variational principle and of the 
role of non-holomorphic corrections it appears to be natural
to contemplate including non-holomorphic terms, thus replacing 
${\cal F}_{\rm OSV}$
by ${\cal F}_{E}$.\footnote{This makes sense microscopically,
because the non-holomorphic corrections to the supergravity
effective action are related to the holomorphic anomaly of the
topological string \cite{BCOV,AGNT}. The role of the holomorphic
anomaly for the OSV conjecture has also been investigated in \cite{VerNH}.} 
Thus we should leave open the option that there are subleading
corrections to the relation between the black hole partition 
function and the topological string partition function.
The weak version of the OSV conjecture \cite{OSV} is:
\be
Z_{\rm mixed}(p, \phi) \approx e^{\pi {\cal F}_{\rm OSV}(p,\phi)} 
= |Z_{\rm top} (p,\phi)|^2 
\;,
\label{MixedZ}
\ee
where $\approx$ means equality in the limit of large charges,
which is the semiclassical limit. Evidence for this form of
the conjecture will be given below. We will also see that
the conjecture needs to be modified as soon as subleading 
corrections are included.

By a formal Laplace transform we can equivalently formulate
this conjecture as a prediction of the state degeneracy 
in terms of the free energy,
by
\be
d(p,q) \approx \int d \phi e^{\pi [ {\cal F}_{\rm OSV} - q \phi]} \;.
\label{StatesOSV}
\ee
Here $d \phi = \prod_I d \phi^I$, and 
the $\phi^I$ are taken to be complex and integrated along  
a contour encircling the origin. The relation (\ref{StatesOSV})
is intriguing,  as it relates the black hole microstates
directly to the topological string partition function.
Note that a saddle point evalation of the integral gives
\be
d(p,q) \approx e^{\Smacro(p,q)} \;,
\ee
because at the critical point of the integrand we have
$\pi [ {\cal F}_E - q_I \phi^I]_{\mscr{attr}} =\Smacro(p,q)$.
Thus the microscopic entropy $\Smicro(p,q) = \log d(p,q)$ and
the macroscopic entropy $\Smacro(p,q)$ argree to leading order 
in the semiclassical limit.\footnote{$\Smacro$ and
$\Smicro$ are expected to be different, once subleading terms
are taken into account, because they refer to different
statisticle ensembles.}

There are several problems which indicate that the proposal  
(\ref{StatesOSV}) must be modified. The number of
states $d(p,q)$ should certainly be invariant under stringy
symmetries such as S-duality and T-duality. In the context of
compactifications with $N\geq 2$ supersymmetry, 
where duality symmetries are 
realized as symplectic transformation, this also means
that $d(p,q)$ should be a symplectic function. However, 
in the approach of \cite{OSV} the electric and magnetic charges
are treated differently, so that there is no manifest 
symplectic covariance. A related issue is how to take into
account non-holomorphic corrections. While \cite{OSV} 
is based on the holomorphic function $F(Y^I, \Upsilon)$, it
is clear that non-holomorphic terms have to enter one way or
another, because they are needed to make
$d(p,q)$ duality invariant. A concrete proposal for 
modifying (\ref{StatesOSV}) was made in \cite{CdWKM:06}.
It is based on the free energy ${\cal F}= 2 H$, i.e., on the
generalized Hesse potential,  instead of ${\cal F}_{\rm OSV}$.
This allows one to treat electric and magnetic charges
on equal footing and to keep symplectic covariance manifest.

The covariant version of (\ref{MixedZ}) is
\be
e^{\pi {\cal F}(\phi, \chi)} = 
e^{2 \pi  H(x,\hat{y})} \approx 
Z_{\rm can}(\phi, \chi) = \sum_{p,q} d(p,q) e^{\pi(q\phi - p \chi)}
\;,
\label{CanZ}
\ee
where $\phi^I  = 2 x^I$ and $\chi_I = 2 \hat{y}_I$ are 
the electrostatic and 
the magnetostatic potentials, respectively,
and $Z_{\rm can}(\phi,\chi)$ is the partition function of the
black hole with respect to the canonical ensemble. 
By a formal Laplace transform we can reformulate the conjecture
(\ref{CanZ}) as a prediction of the state degeneracy:
\be
d(p,q) \approx \int d x  d \hat{y} 
e^{\pi \Sigma(x, \hat{y},p,q)} \;.
\label{StatesUS}
\ee
In absence of $R^2$- and non-holomorphic 
corrections, the measure 
$dx d y = \prod_{I,J} dx^I d y_J$
is proportional to the top power
of the symplectic form $dx^I \wedge dy_I$ on 
$C(M_{VM})$ and therefore is symplectically invariant. 
In the presence of $R^2$- and non-holomorphic corrections, 
$dx d\hat{y}$ is the appropriate generalization.
Since $\Sigma$ is a symplectic function, we have found
a manifestly symplectically covariant version of
(\ref{StatesOSV}).

As before, the variational principle ensures that in 
saddle point approximation we have 
$d(p,q) \approx \exp(\Smacro)$, as $\Smacro$ is the
Legendre transform of the Hesse potential and hence the
saddle point value of $\pi \Sigma$. In order to compare
(\ref{StatesUS}) to (\ref{StatesOSV}), we can rewrite
(\ref{StatesUS}) in terms of the complex variables and
perform the integral over $\mbox{Im}Y^I$ in saddle 
point approximation, i.e., we perform a Gaussian
integration with respect to the subspace where the
magnetic attractor equations are satisfied. The result
is \cite{CdWKM:06}
\be
d(p,q) \approx \int d \phi \sqrt{\Delta^- (p,\phi) }
e^{\pi [ {\cal F}_E - q \phi ]}
\ee
and modifies (\ref{StatesOSV}) in two ways: first, 
in contrast to \cite{OSV} we have 
included  non-holomorphic terms into the free energy
${\cal F}_E$; second, the integral contains
a measure factor $\Delta^- (p,\phi)$, whose explicit form
can be found in \cite{CdWKM:06}. The measure factor 
is needed in order to be consistent with symplectic covariance.

The proposals (\ref{StatesOSV}) and (\ref{StatesUS})
can be tested by comparing
the black hole entropy to the microscopic state degeneracy.
There are some cases where these are either known exactly,
or where at least subleading contributions are accessible. 
While this chapter is far from being closed, there seems
to be agreement by now that (\ref{StatesOSV})
needs to be modified by a measure factor \cite{DDMP,ShiYin,CdWKM:06}. 
In particular, the measure factors extracted from the evaluation of 
exact dyonic state degeneracies in $N=4$ compactifications
\cite{Sen} 
are consistent, at the semiclassical level, with the
proposal (\ref{StatesUS}) \cite{CdWKM:06}. Detailed investigations
of microscopical $N=2$ partition functions have clarified the
origin of the asymptotic holomorphic factorization of the
black hole partition function, $Z_{\rm mixed} \approx |Z_{\rm top}|^2$:
it results from simultanous contributions of branes and anti-branes
to the state degeneracy \cite{GSY,KL,BGGHSY,BCDMV}. 
Recently, the refined analysis of \cite{DenMoo} has identified
a microscopic measure factor, which agrees with the one
found in \cite{ShiYin,CdWKM:06} in the semiclassical limit.

\subsection*{Acknowledgments}

The original results reviewed in this article were obtained in 
collaboration with Gabriel Lopes Cardoso, Vicente Cort\'es, 
Bernard de Wit, J\"urg K\"appeli, 
Christoph Mayer and Frank Saueressig. The article
is partially based on a talk given at the `Bernardfest' in Utrecht,
and I would like to thank the organisers for the opportunity to 
speak on the occasion of Bernard's anniversary. My special thanks
goes to Vicente Cort\'es 
for inviting me to contribute this article to the 
Handbook on Pseudo-Riemannian Geometry and Supersymmetry.
Furthermore, I would like to thank the referee for suggesting 
to make this article accessible to a larger readership 
by including an introduction 
to black holes in general  and to supersymmetric black 
hole solutions in particular.
This material has been incluced as a separate section
(Section \ref{BHsandStrings}). The extended version of this article was
written during a stay as Senior Research Fellow
at the Erwin Schr\"odinger for Mathematical Physics in Vienna.

\end{document}